# A New Approach to Proportional Hazards Modeling for Estimating Customer Lifetime Value

Vadim Pliner
Afiniti
vadim.pliner@afiniti.com

Estimating customer lifetime value (CLV or LTV) is extremely important for making better business decisions. The proposed flexible proportional hazards model allows an estimation of lifetime value in contractual settings. This approach takes advantage of a churn model, which is assumed to be available.

1. Introduction.

According to Wikipedia, "in marketing, customer lifetime value (CLV or often CLTV), lifetime customer value (LCV), or life-time value (LTV) is a prognostication of the net profit contributed to the whole future relationship with a customer." It's a forward looking measure of both customer profitability and loyalty. A recent paper in Harvard Business Review [1] says: "given the importance of customer value, leaders should track it as rigorously as they track other key assets, such as buildings, machinery, inventory, and marketable securities". Our paper deals with contractual or subscription business settings where the customer churn/attrition/turnover is observed. Examples include telecommunication, banking, and insurance industries. Mathematically speaking,

$$CLV = \sum_{t=1}^{\infty} p_t \times M_t \times (1+r)^{-t}, \qquad (1)$$

where
- $p_t$ is the estimated probability to remain the company's customer for the next t months or weeks or any specified *t* time periods,
- $M_t$ is the expected profit margin (revenue minus costs) during the period t, and
- r is the discount rate, which is the interest rate used to determine the present value of future cash flows. It accounts for the way the value of money is discounted over time. The concept is based on the premise that a dollar today is worth more than a dollar tomorrow. Discounting may or may not be taken into account.

Data on the individual customer costs may not be readily available. Then revenues can be used instead of margins in (1). The future margins can be either predicted (e.g. see [2]) or assumed to be constant and equal to either the latest customer's margin or its average over several most recent time periods. In case when $M_t$ are not predicted and discounting is not applied, the formula for CLV above can be simplified to

$$CLV = M \times \sum_{t=1}^{\infty} p_t \tag{2}$$

In this paper we'll be dealing with the "survival" part of CLV and will propose a model for estimating the survival probabilities $p_t$. We assume there is a churn model in place, allowing to estimate (at the customer level) the probability of churn over the next month or another typically short period of time. The proposed CLV approach will heavily rely on that churn model.

Another important and related to CLV metric is the remaining tenure (RT). The mathematical expectation of RT is known to be:

$$E(RT) = \sum_{t=1}^{\infty} p_t, \tag{3}$$

So, if the time unit $t$ is month, for example, the quantity in (3) will be the mean expected remaining tenure in months. We can then reformulate (2) as

$$CLV = M \times E(RT).$$

From now on we'll assume the time unit is month keeping in mind it can be any time unit. For practical reasons, the time horizon is typically limited by a fixed number of months and the summation in (1) and (2) is done up to that number rather than to infinity.

Estimating $p_t$ lies in the scope of survival analysis, also known as failure time analysis and event history analysis. Two main functions in survival analysis are the hazard function $h(t) = P(T = t \mid T \geq t)$, that is the probability that an event (in our case customer churn) occurs during time interval $t$ and the survival function, which is $S(t) = P(T > t)$. In fact, the values of $p_t$ in (1) – (3) equal to $S(t)$. Estimating only one of the two functions is sufficient, because

$$S(t) = S(t-1) \times [1-h(t)] \tag{4}$$

and therefore,

$$S(t) = [1-h(1)] \times [1-h(2)] \times \ldots \times [1-h(t)] \tag{5}$$

Below we propose a proportional hazards model for estimating values of the hazard function. It's different from the well-known Cox proportional hazards model aka. the Cox regression [3] and we'll discuss the differences later.

The probability to churn and CLV are closely related in the same way the hazard and survival functions are related. However, churn models typically predict churn over a much shorter horizon (and they are good at that) than CLV models predict survival. Our approach tries to take advantage of the predictive power of churn models to sensibly extend their predictions to longer periods of time.

## 2. Proportional hazards model

The proposed model stipulates that for each customer *i*, his/her hazard function

$$h_i(t) = \alpha_i \times h_0(t), \quad (6)$$

where the baseline hazard function $h_0(t)$ is the hazard function over the whole customer base and $\alpha_i$ is the coefficient of proportionality, which is unique by customer.

The baseline hazard function $h_0(t)$ can be estimated using historical data with either standard survival analysis techniques, e.g. the Kaplan-Meier method for estimating the survival function [4] and using (4) to get to the hazard function or simply computing monthly churn rates by tenure. The application of the Kaplan-Meier method requires historical data going back quite a distance in time. However, if we take a snapshot of our data as of at least a month ago, we can compute hazards (churn rates) by tenure. In fact, a snapshot is too strong of a word, all we need are customer tenures at some point in time and the churn status over a month after that point for each customer.

We should also provide estimates of the baseline hazard function beyond the scope of tenures in historical data as going forward the existing customers' tenures will increase. Typically, the hazard function stabilizes beyond some point. To make sure this is the case, it is a good idea to plot the baseline hazard function to see such a point and compute the mean hazard/churn rate for tenures beyond that point. Then we can extrapolate the hazard function values using that mean for tenures higher than seen in historical data.

Now, let $t_0$ be the current tenure of customer i. Then $h_i(t_0)$ is the probability to churn over the next month and can be estimated with the churn model, the existence of which we assumed in the introduction. Therefore, $\alpha_i$ in (6) can be estimated as $h_i(t_0)/h_0(t_0)$ and for all $t \geq t_0$,

$$h_i(t) = h_0(t) \times h_i(t_0)/h_0(t_0) \quad (7)$$

It is theoretically possible that some values of $h_i(t)$ in (7) are greater than 1. In those rare cases, we will set $h_i(t)$ to 1.

A graphical illustration of the approach is shown on the plot below based on fictitious data. In blue here is a baseline hazard function, the current tenure of customer *i* is 18 months, and $\alpha_i$ is 1.3, meaning that the probability to churn over the next month (based on the churn model) is 1.3 times higher than the churn rate among all customers with 18 months of tenure.

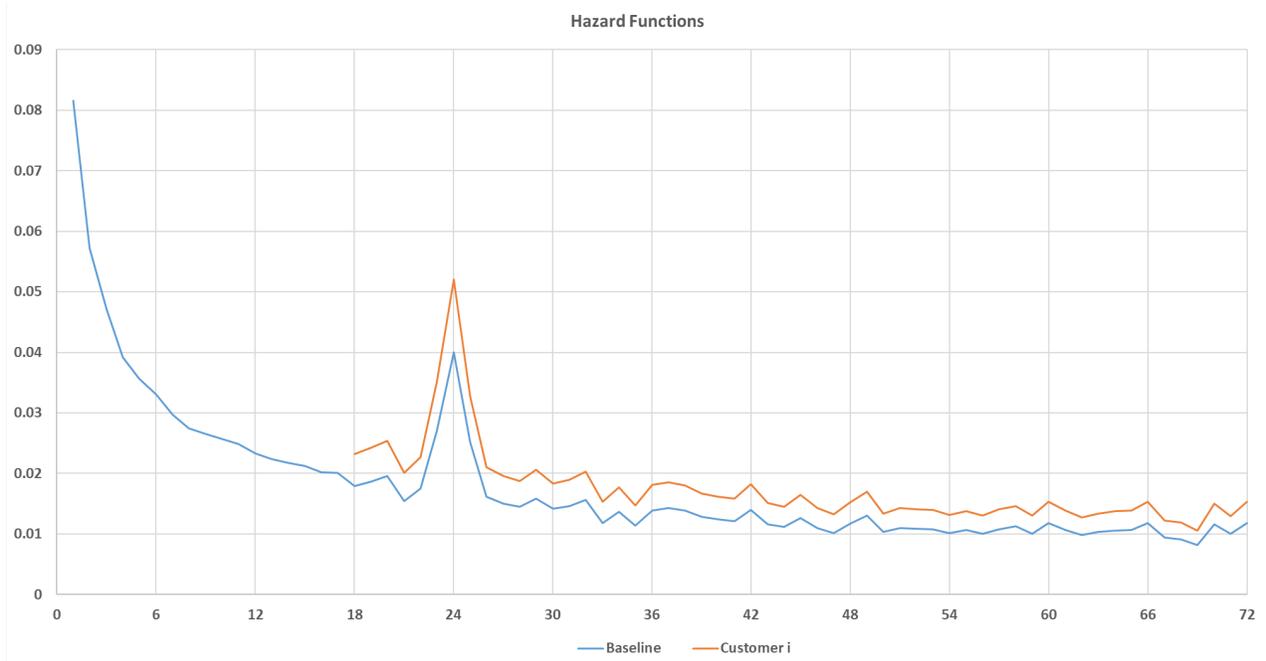

After getting estimates of the hazard function, the survival function can be estimated with (5), that is for each customer *i* and tenure $t > t_0$, where $t_0$ is the current tenure,

$$S_i(t) = \prod_{j=t_0}^{t} [1-h_i(j)]$$

And the mathematical expectation of remaining tenure of customer *i* is

$$E(RT_i) = \sum_{j=0}^{\infty} S_i(t_0+j) \tag{8}$$

We do not want to do summation in (8) to infinity of course. Luckily, the sum's incremental values become increasingly smaller and one can terminate summation when they are deemed sufficiently small.

When there are two separate churn models, for voluntary and involuntary churn (usually for non-payment reasons), the probability to churn can obviously be estimated as the sum of the two models' scores. However, a more granular approach is to treat the two mutually exclusive events (voluntary and involuntary churn) as "competing risks." Then one would define two sub-hazard functions:

$$h_v(t) = P(T_v = t \mid T \geq t) \text{ and } h_{inv}(t) = P(T_{inv} = t \mid T \geq t),$$

for voluntary and involuntary churn, respectively, and apply the above approach to the two functions separately. In this scenario, the overall hazard function

$$h(t) = h_v(t) + h_{inv}(t)$$

3. Similar traditional approaches to CLV modeling

Our model is not the only proportional hazard model in existence. There is of course the well-known Cox proportional hazards model aka. the Cox regression [3]. It is quite different though.

The customer's hazard function according to the Cox model can be expressed as

$$h(t) = h_0(t) \times \exp\left(\sum_{j=1}^{m} \beta_j x_j\right) \quad (9)$$

where $x_j$ are covariates/predictors and $h_0(t)$ is the baseline hazard function. It's not the same baseline function that we discussed. Every combination of predictors' values defines a specific hazard function. $h_0(t)$ obviously corresponds to the case when all of them are zeroes.

The coefficient of proportionality is a function of static predictors. The use of time-dependent covariates in the context of CLV is feasible only if we predict their future values. The model's parametric functional form is very specific. The Cox model was not designed with CLV in mind, but rather for biomedical applications with the main focus on beta coefficients in (9), which in fact can be estimated without knowing or estimating $h_0(t)$.

Also used in the context of CLV, the discrete-time logistic hazard models were first introduced in biostatistics [5], but found more demand in the social sciences [6] and business applications [2,7]. They are defined in the following way:

$$\text{Log}\{h(t)/[1-h(t)]\} = \psi(t,\alpha) + \sum_{j=1}^{m} \beta_j x_j(t), \quad (10)$$

where $\psi(t,\alpha)$ is a cubic spline and $x_j(t)$ are covariates/predictors that also are typically static rather than functions of $t$ in the context of CLV. Or, in terms of odds,

$$h(t)/[1-h(t)] = \exp[\psi(t,\alpha)] \times \exp\left(\sum_{j=1}^{m} \beta_j x_j\right) \quad (11)$$

If we set all $x_j$ to 0, we can see from (11) that the model is trying to approximate the odds, i.e. $h(t)/[1-h(t)]$ using the exponential function of cubic splines. Cubic splines is a pretty flexible class of functions, but we don't think this approximation is necessary at all. In case when there are enough observations with $x_j = 0$ for $j=1,\ldots, m$, $h(t)/[1-h(t)]$ can be directly and non-parametrically estimated similarly to the described above ways to estimate the baseline hazard function instead of approximating it with $\exp[\psi(t,\alpha)]$. In other words, we can replace (11) with

$$h(t)/[1-h(t)] = h_0(t)/[1-h_0(t)] \times \exp\left(\sum_{j=1}^{m} \beta_j x_j\right), \quad (12)$$

where $h_0(t)/[1-h_0(t)]$ is the baseline hazard odds function, and the model becomes very similar to the Cox regression. (12) is also a proportional model, but instead of hazards, hazard odds are proportional. Whether there are enough or not enough observations with all $x_j = 0$, we can suggest using $h_0(t)$ based on the whole customer base of interest as in our approach offered above.

Going back to the general form of the discrete-time logistic hazard model (10), its suggested modification looks like this:

$$\text{Log}\{h(t)/[1-h(t)]\} = \log\{h_0(t)/[1-h_0(t)]\} + \sum_{j=1}^{m} \beta_j x_j(t),$$

where the baseline hazard function $h_0(t)$ can be estimated with computing monthly churn rates by tenure as was proposed in 2. above.

In summary, we introduced a new approach to CLV estimation. It is based on a proportional hazards model, which we believe is more flexible and powerful than similar traditional methods, although the existence of a churn model is assumed.